\documentclass[a4paper,fleqn,usenatbib]{mnras}
\usepackage[T2A, T1]{fontenc}
\usepackage{ae,aecompl}
\usepackage{graphicx}
\usepackage{amsmath}
\usepackage{amssymb}
\usepackage{array,tabularx}
\usepackage{ulem}
\usepackage[english]{babel}
\usepackage{threeparttable}
\usepackage{enumitem}
\usepackage{xcolor,cancel}
\makeatletter
\let\TPT@hookin\@gobble
\let\TPT@hookarg\@gobble
\makeatother

\title[AX J1745.6-2901: evolution]{Population synthesis of AX J1745.6-2901 X-ray nova type binaries with rapidly decreasing orbital periods}

\author[A. I. Bogomazov et al.]{A. I. Bogomazov$^1$\thanks{E-mail:a78b@yandex.ru (AIB)},
A. M. Cherepashchuk$^1$\thanks{E-mail:Cherepashchuk@gmail.com (AMC)},
T. S. Khruzina$^1$\thanks{E-mail:kts@sai.msu.ru (TSK)},
and A. V. Tutukov$^2$\thanks{E-mail:atutukov@inasan.ru (AVT)}
\\
$^1$M. V. Lomonosov Moscow State University, P. K. Sternberg Astronomical Institute, 119234, Universitetkij prospect, 13, Moscow, Russia\\
$^2$Institute of Astronomy, Russian Academy of Sciences, 119017, Pyatnitskaya st., 48, Moscow, Russia\\
}

\date{Accepted . Received ; in original form }

\pubyear{2021}

\begin{document}
\label{firstpage}
\pagerange{\pageref{firstpage}--\pageref{lastpage}}
\maketitle

\begin{abstract}
The neutron star low mass X-ray binary (LMXB) AX J1745.6-2901 was detected an anomalously fast decrease of orbital period. The decreasing rate of orbital period exceeds the contribution of all processes extracting angular momentum from the binary star in standard model. Using {\ttfamily SCENARIO MACHINE} code we conducted a population synthesis study of X-ray novae with neutron stars to investigate a possible formation and evolution of such binaries. Such close LMXBs should experience a preceding common envelope stage, in which the magnetic fields of the low mass main-sequence donor stars can be dramatically amplified. Our calculations show that the magnetic stellar wind of the optical companion can efficiently extract angular momentum from the binary systems, and produce the observed orbital-period derivatives of AX J1745.6-2901 and black hole LMXBs. The estimated values of the required magnetic field induction are following: $B_\textrm{d}\approx 400$ G (AX~J1745.6-2901), $B_\textrm{d}\approx 1500$ G (KV UMa), $B_\textrm{d}\approx 400$ G (A0620-00), $B_\textrm{d}\approx 1800$ G (Nova Muscae). We successfully reproduced the current observational abundance of such anomalous neutron star X-ray novae, and computed the appropriate value of the parameter of magnetic braking $\lambda_\textrm{MSW}$ ($0.8-0.6$ for Roche lobe filling stars and $0.4-0.15$ for binaries with partial Roche lobe filling).
\end{abstract}


\section{Introduction}
\label{sec1}

An anomalously rapid decrease of the orbital period of several X-ray novae with black holes (KV UMa, A0620-00 and Nova Muscae 1991) in quiescence was discovered by \citet{g-h-2012ApJ,g-h-2014,g-h-2017}. \citet{ponti2017} found the same anomalously fast decrease of the orbital period of a low mass X-ray system with a neutron star AX~J1745.6-2901.

An evolutionary status of X-ray binaries with black holes and a possible cause of the fast decrease of the orbital period was studied by \citet{cherepashchuk2019}. They found that the anomalous decrease of the orbital period in KV UMa, A0620-00 and Nova Muscae 1991 can be explained by the highly increased angular momentum loss via the magnetic stellar wind of the non-degenerate star. The magnetic field of the optical star could be highly amplified during preceding common envelope stage (see appropriate calculations by \citealp{ohlmann2016}). The anomaly in mentioned black hole binaries potentially can be explained by the interaction between the binary and a surrounding circumbinary disc \citep{chen2015,xu2018,chen2019}. These mechanisms can work together with other mechanisms that can change the angular momentum of the system, they can have different efficiency that can be a matter of debates.

A potential role of the change of the gravitational momentum of the magnetized optical star \citep{applegate1992} as the explanation of the fast rate of the orbital shrinkage in AX~J1745.6-2901 was discussed by \citet{ponti2017}. They came to a conclusion that this mechanism cannot explain the anomaly, because the magnetic activity should change the duration of eclipses and/or the ingress or egress time in an uncorrelated way, whereas these features do not take place in the system. A non-conservative mass transfer and a possible gravitational influence of a third body in the system were considered as potential explanations of the anomaly \citep{ponti2017}.

The aim of the present study is to investigate a potential role of the increased angular momentum loss via the magnetic stellar wind of the optical star in case of the AX~J1745.6-2901 system and to study the evolutionary path of such X-ray novae.

\section{On the birth frequency of low mass X-ray binaries and on their quantity in the Galaxy}

\citet{kiel2006} calculated the birth rate of low mass X-ray binaries (LMXBs) in the Galaxy as $\approx 4\times 10^{-7}$ yr$^{-1}$ for black hole binaries and $\approx 6\times 10^{-7}$ yr$^{-1}$ for neutron star binaries to meet the observed quantity of LMXBs $\sim 10^3$ in the Galaxy. Let us make a simple and semi-qualitative description of the origin of such estimations.

The birth rate of binary stars in the Galaxy is \citep{tutukov1988}:

\begin{equation}
\partial^3\nu \approx 0.2 \partial\log a_0 \frac{\partial M_{01}}{M_{01}^{2.35}}f(q)\partial q\, \mbox{yr}^{-1},
\label{tutukov-equation-1}
\end{equation}

\noindent where $a_0$ is the initial separation of stars in units of $R_\odot$ ($10R_\odot\leq a_0\leq 10^6R_\odot$), $M_{01}$ is the initial mass of the primary in units of $M_\odot$ ($0.1M_\odot\leq M_{01}\leq 150M_\odot$), $q=M_{02}/M_{01}$ ($0<q\leq 1$) is the initial mass ratio of the components, $f(q)$ is the initial distribution of binaries on $q$.

The mass function \citep{salpeter1955,kroupa2019} gives following birth frequency for neutron stars (the initial mass of the progenitor is $10M_{\odot}-25 M_\odot$) $\nu\simeq \int\limits_{10M_\odot}^{25M_\odot}\dfrac{dM_{01}}{M_{01}^{2.35}}\simeq 2.5\times 10^{-2}$ yr$^{-1}$, for black holes (the initial mass of the progenitor is $\geq 25 M_\odot$) the frequency is $\nu\simeq \int\limits_{25M_\odot}^{150M_\odot}\dfrac{dM_{01}}{M_{01}^{2.35}}\approx 9\times 10^{-3}$ yr$^{-1}$. For these estimations dimensionless integrals $0.2\int\limits_{10R_\odot}^{10^6R_\odot} d\log a_0$ and $\int\limits_{0}^{1} f(q)dq$ both are equal to 1, the whole range of separation and mass ratio was taken into account, $f(q)=1$ (equiprobable mass ratio). The latter upper limit of stellar masses weakly affects the frequency due to the power dependence of the frequency on the mass, it can be taken very different according to variety of suggestions. The uncertainty of such estimations is within the factor 2-3, for example, the part of binary stars among all stars can be less than 100\% (these estimations are made for the assumption that all stars are binaries).

The neutron stars and stellar-mass black holes are usually believed to be results of supernova explosions, so the above estimate can be compared with the rate of supernova explosions. The supernova rate in the Milky Way galaxy was estimated as one explosion in $40\pm 10$ yr (see, e.g., \citealp{tammann1994}), \citet{cappellaro1997} found that faint events can potentially increase the estimated rate of supernovae by no more than 20-30\%. The recent estimation of core collapse supernovae rate in our galaxy made by \citet{rozwadowska2021} is $1.63\pm 0.43$ per century. It is in excellent agreement with the result obtained using Equation \ref{tutukov-equation-1} taking into account the uncertainty of the part of binaries among all stars.

The distribution on the initial semi-major axis gives the frequency that depends on the efficiency of the common envelope. For $\alpha_\textrm{CE}\simeq 1$ (see, e.g., Equation A3 by \citealp{cherepashchuk2019}) the maximum semi-major axis for the progenitors of LMXBs with an ordinary magnetic stellar wind ($\lambda_\textrm{MSW}=1$, Equations A4 and A5 by \citealp{cherepashchuk2019}) of the main sequence donor star (Equation A5 by \citealp{cherepashchuk2019}, the maximum age is the age of the Universe that is approximately equal to $13\,900$ millions of years, and the radius of the donor star is taken to be $R_\textrm{d}/R_\odot=M_\textrm{d}/M_\odot$) with the mass in the range $0.3 M_{\odot}$-$1.5 M_{\odot}$ is $a_\textrm{max}/R_\odot\approx 5\left(M_\textrm{a}/M_\odot\right)^{1/5}\left(M_\textrm{d}/M_\odot\right)^{4/5}$, where $M_\textrm{d}$ and $M_\textrm{a}$ are the masses of the accretor and donor correspondingly. The minimum semi-major axis is determined by the condition of the full filling of the Roche lobe by the donor (see Equation \ref{a-roche}): $a_\textrm{min}/R_\odot\approx 2.16 \left(M_\textrm{a}/M_\odot\right)^{1/3}\left(M_\textrm{d}/M_\odot\right)^{2/3}$. The ratio of these quantities is $a_\textrm{max}/a_\textrm{min}\approx 2(M_\textrm{d}/M_\textrm{a})^{2/15}$. Therefore, the distribution on the initial semi-major axis gives the factor $f(a)\approx 0.06$ in Equation (\ref{tutukov-equation-1}).

The initial distribution of components on the mass ratio is very uncertain. \citet{kraicheva1981,gullikson2016} give $dN/dq=1$, and the change of this quantity for stars with the mass $M_{01}\leq 1M_\odot$ can be up to the factor $\approx 5$ \citep{lineweaver2005}. Assuming $dN/dq=1$ for LMXBs with neutron stars the value is $f(q)\approx 0.05$ and for systems with black holes it is $f(q)\approx 0.02$.

Taking into account the contribution of all factors in Equation (\ref{tutukov-equation-1}) we can find the birth frequency of neutron star LMXBs as $\sim 10^{-4}$ yr$^{-1}$, for the black hole binaries it is $\sim 10^{-5}$ yr$^{-1}$. The uncertainty of these estimates is about an order (or even several orders) of magnitude, and they can be significantly higher than the empirical value.

The birth rates of neutron stars and black holes computed using the initial Salpeter mass function coincide with the observational rate of core collapse supernovae within the factor of a few, it is the simplest part of frequency estimations. Complicated evolution of close binary stars and strong observational selection effects make all observational estimations of quantities and birth frequencies of LMXBs very uncertain. A naive example of an estimation can be following. The production of the lifetime of an ordinary X-ray nova and the birth frequency of such novae in the Galaxy gives the number such systems in the Galaxy. The number of X-ray novae (both neutron star and black hole binaries) in our galaxy is $\sim 10^3$ \citep{chen1997,cherepashchuk2013}, if the assumed lifetime of such binary equals to the solar hydrogen burning time ($\sim 10^{10}$ yr) then the birth frequency is $\sim 10^{-7}$ yr$^{-1}$. Usually estimations of birth frequencies and quantities of such objects as X-ray novae include results of binary population synthesis models (see, e.g., \citealp{kiel2006}, their frequency is $\sim 10^{-6}$ yr$^{-1}$, or \citet{pfahl2003}, their frequency is $\sim 10^{-6}-10^{-4}$ yr$^{-1}$).
 
The rest discrepancy in principle can be explained by the uncertainty of the initial mass ratio of binaries $dN/dq\ll 1$. But also it is possible to treat LMXBs as non-conservative systems due to the induced stellar wind of the donor heated by the X-ray radiation from the accretor. Let us consider such situation (see also \citealp{iben1997,tutukov2002} for the study about black hole binaries).

The conditions of the existence of the induced stellar wind in LMXBs are following. Each gram of the matter accreted by the relativistic object emits $\sim 10^{20}$ erg. A part of this energy (in case of a spherical symmetry) can be absorbed by the donor. This value is $\left(R_\textrm{d}/a\right)^2\times 0.25\times 10^{20}$ erg. If the energy can be ``reprocessed'' with the $\alpha$ efficiency to the donor's stellar wind (its binding energy is $\sim 10^{15}$ erg g$^{-1}$) this energy leads to the donor's mass loss $\approx 2.5\times 10^4\alpha\left(R_\textrm{d}/a\right)^2$ g. As the result the accretion of one gram of matter leads to the donor's loss about $4\times 10^3\alpha \left(M_\textrm{d}/(M_\textrm{a}+M_\textrm{d})\right)^{3/2}$ g. The same value of the gas loss can be obtained in the supposition that the evaporation takes place on the outer part the accretion disc instead of the donor star. It is clear that if $\alpha\approx 1$ the non-conservativeness (due to the induced stellar wind of the donor and/or disc) makes it possible to decrease the birth frequency of LMXBs by 2 or 3 orders of magnitude and to meet the required frequency $\sim 10^{-6}$  yr$^{-1}$ in the Galaxy and even lower than it. This discussion sharpens the issue about the effectiveness of the generation of the induced stellar, this issue requires the development of an adequate gas dynamic model.

Also it is useful to show that the effectiveness of the accretion of the donor's wind should be low. For $M_\textrm{a}$, $M_\textrm{d}$, $R_\textrm{d}$ (in the supposition that the velocity of matter in the wind is $\sim \sqrt{2GM_\textrm{d}/R_\textrm{d}}$) the part of the wind intercepted by the accretor should be: $\approx (1/4)(R_\textrm{d}M_\textrm{a})/(aM_\textrm{d})^2\approx 0.01(M_\textrm{a}/M_\textrm{d})^{4/3}$. For neutron stars as accretors for $M_\textrm{d}\approx 0.5M_\odot$ the effectiveness of the accretion of the donor wind's matter becomes $\approx 0.04$, and for black holes with mass $\approx 5 M_\odot$ the effectiveness becomes $\approx 0.04$, and for black holes with mass $\approx 5 M_\odot$ and for the donor's mass $M_\textrm{d}=0.5 M_\odot$ it should be $\approx 0.2$. For the more reliable estimate of the effectiveness of the accretion there is again a need to develop a detailed gas dynamic model, the described simplified analysis showed a potential importance of the non-conservativeness of LMXBs for estimates of their birth frequencies using the population synthesis method.

However, for example, \citet{avakyan2021} considered a mechanism for the removal of angular momentum from an X-ray binary system and the change in its orbital period (the mass loss in the form of a wind from an accretion disc) and came to a conclusion that outbursts in windy accretion discs are not able to explain the mentioned above rapid decrease of the orbital period in KV UMa, A0620-00 and Nova Muscae 1991. Therefore in this paper we study a potential role of the increased angular momentum loss via the magnetic stellar wind as in the work by \citet{cherepashchuk2019}.

\section{On the angular momentum loss after the common envelope}

Let us make a simple semi-qualitative estimation of the angular momentum loss via the magnetic stellar wind after the common envelope phase.

For the frozen magnetic field the magnetic flux of the spherical stellar wind can be expressed as $BR^2=C$, where $B$ is the magnetic induction, $R$ is the distance from the optical star, $C$ is a constant value. For the magnetic energy density it gives

\begin{equation}
\varepsilon_\textrm{B}= \frac{B^2}{8\pi}= \frac{C^2}{8\pi R^4}.
\label{magnetic-energy}
\end{equation}

The wind's energy density is

\begin{equation}
\varepsilon_\textrm{w}= \frac{1}{2}\rho w^2= \frac{1}{2} \frac{\dot M_d}{w4\pi R^2} w^2= \frac{1}{8\pi}\frac{\dot M_d w}{R^2},
\label{wind-energy}
\end{equation}

\noindent where $\dot M_d$ is the mass loss rate via the stellar wind, $\rho$ is the density of the wind, $w$ is its velocity.

The equality of these energies can be established for the radius

\begin{equation}
R_\textrm{w}=\left( \frac{C^2}{\dot M_d w} \right)^{1/2};
\label{raduis-momentum}
\end{equation}

\noindent within $R_\textrm{w}$ from the star the wind is ``embedded'' into the magnetic field lines, outside it the wind breaks off the lines and freely flows outwards the optical star.

The angular momentum of the system is

\begin{equation}
J=\frac{1}{(2\pi)^{1/3}}\frac{M_\textrm{a} M_\textrm{d}}{(M_\textrm{a} + M_\textrm{d})^{1/3}}G^{2/3}P_\textrm{orb}^{1/3},
\label{angular-momentum}
\end{equation}

\noindent where $M_a$ is the mass of the accretor, $M_d$ is the mass of the donor, $P_{orb}$ is the orbital period, $G$ is the gravitational constant.

The rate of the angular momentum loss due to the magnetic stellar wind is

\begin{equation}
\dot J = - \dot M_d R^2_\textrm{w}\omega = - \frac{\dot M_d \omega C^2}{\dot M_d w} = - \frac{\omega C^2}{w},
\label{momentum-loss}
\end{equation}

\noindent where $\omega$ is the angular velocity of the optical star's axial rotation. The tidal forces between the neutron star and the donor star would continuously spin the donor star back up into co-rotation with the orbital rotation. The spin-up takes place at the expense of the orbital angular momentum. Hence the magnetic stellar wind indirectly carries away the orbital angular momentum of the binary system.

The orbital period derivative obtained from Equations (\ref{angular-momentum}), (\ref{momentum-loss}) and from the Kepler's third law is 

\begin{equation}
\dot P_\textrm{orb} = - 3(2\pi)^{4/3}\frac{(M_\textrm{a}+M_\textrm{d})^{1/3}}{M_\textrm{a}M_\textrm{d}}\frac{1}{G^{2/3}}\frac{C^2}{w}\frac{1}{P_\textrm{orb}^{1/3}}.
\label{period-derivative}
\end{equation}

Equation (\ref{period-derivative}) does not consider the orbital angular momentum loss due to the donor wind. Since the wind mass loss rate by a low mass star should be very low ($\lesssim 10^{-13}M_\odot$ yr$^{-1}$, see, e.g., \citealp{wood2002}), such loss can be neglected to make estimations of the magnetic field induction\footnote{The {\ttfamily SCENARIO MACHINE} code takes into consideration the orbital angular momentum loss because of the mass loss via the stellar wind.}.

To explain the increased angular momentum loss $\dot J$ following suggestions can be considered:

\begin{enumerate}

\item The greater size of the donor leads to the less $w$.

\item The stronger magnetic field of the donor leads to the greater $C$.

\item The greater the angular velocity of the donor leads to the greater $\dot J$ even for the same size and the same magnetic field strength.

\end{enumerate}

Let us consider items 1 and 3. Possible differences of the radius of the donor star in a close low mass X-ray binary cannot be too high, so the difference of $w$ also seems to be small. The equatorial rotational velocity of the donor star that fills its Roche lobe and rotates synchronously with the orbital movement from the Kepler's third law is

\begin{equation}
v_\textrm{e} = R_\textrm{d}\sqrt{\frac{G(M_\textrm{a}+M_\textrm{d})}{a^3}},
\label{ve}
\end{equation}

\noindent where $R_\textrm{d}$ is the radius of the optical star.

The maximum possible velocity on the surface of the star is Kepler velocity for this star

\begin{equation}
v_\textrm{max}=\sqrt{\frac{GM_\textrm{d}}{R_\textrm{d}}}.
\label{vmax}
\end{equation}

The major semi-axis and radius of the Roche lobe filling optical star are connected by the following formula\footnote{This formula can be applied for binaries with $M_\textrm{d}\leq 0.8 M_\textrm{a}$, it is valid for X-ray novae considered in this work and for most X-ray novae taking into account the estimated values of masses of donors in them ($<1.1M_\odot$, mostly less and much less that the solar mass, \citealp{cherepashchuk2013}). SCENARIO MACHINE code uses more careful fit using the formula by \citet{eggleton1983} that is valid for any value of the mass ratio in the binary.} \citep{paczynski1971}:

\begin{equation}
R_\textrm{d}=0.46(M_\textrm{d}/(M_\textrm{a}+M_\textrm{d}))^{1/3}a.
\label{a-roche}
\end{equation}

The ratio of the maximum angular velocity $\omega_\textrm{max}$ and the angular velocity $\omega$ that is equal to the orbital angular velocity is

\begin{equation}
\frac{\omega_\textrm{max}}{\omega}= \frac{v_\textrm{max}}{v_\textrm{e}}=4;
\label{omega-ratio}
\end{equation}

\noindent i.e., the axial rotation of the donor star potentially can be accelerated in the common envelope phase by four times.

If $v_\textrm{e}=\lambda_\textrm{MSW}\times 10^{14} \times t^{-1/2}$ cm s$^{-1}$\citep{skumanich1972} then:

\begin{eqnarray}
\frac{dJ}{dt}=\frac{d(I_\textrm{d}\Omega_\textrm{d})}{dt}&=&\frac{\lambda_\textrm{MSW}10^{14}I_\textrm{d}}{R_\textrm{d}}\frac{dv_\textrm{e}}{dt}=\nonumber \\
=-\frac{\lambda_\textrm{MSW} 10^{14}I_\textrm{d}}{2R_\textrm{d}} \frac{1}{t^{3/2}}&=&\nonumber \\
=-\frac{I_\textrm{d}}{2\lambda_\textrm{MSW}^2\times 10^{28}R_\textrm{d}} v_\textrm{e}^3;
\label{axil-acceleration}
\end{eqnarray}

\noindent where $t$ is the age of the optical star, $\Omega_\textrm{d}$ is the angular velocity of the donor's axial rotation, $I_\textrm{d}$ is the moment of inertia of the donor; i.e., the angular momentum loss can be accelerated by a factor of 64. This quantity is greater than the discrepancy between the usual estimation of the magnetic braking and the braking force required to explain the accelerated decrease of the orbital period in KV UMa, A0620-00, Nova Muscae 1991 and AX~J1745.6-2901.

However, the axial rotation of the Roche lobe filling donor star with the rate faster than the orbital rate potentially can be only in a relatively short time scale after the end of the common envelope phase, than the axial rotation should be practically synchronized with orbital rotation due to severe tides in such close system with the Roche lobe filling donor star.

Therefore the main idea to explain the accelerated shrinkage of the system under consideration in this paper is the amplification of the magnetic field during the common envelope.

The value of $B_d$ on the surface of the donor can be calculated from Equation (\ref{period-derivative}) (taking into account that $C=BR^2$):

\begin{equation}
B_d=\frac{|\dot P_\textrm{orb}|^{1/2}}{3^{1/2}(2\pi)^{2/3}}\frac{(M_\textrm{a}M_\textrm{d})^{1/2}}{(M_\textrm{a}+M_\textrm{d})^{1/6}}G^{1/3}\frac{w^{1/2}}{R_\textrm{d}^2}P_\textrm{orb}^{1/6}.
\label{induction}
\end{equation}

Estimations of the magnetic induction of considered binaries required to explain the accelerated shrinkage of their orbits can be found in Section \ref{magnetic-estimations}.

\section{The Scenario Machine}
\label{sec:scm}

In this paper we continued the investigation by \citet{cherepashchuk2019} and studied the evolutionary status of low mass X-ray binaries with neutron stars using {\ttfamily SCENARIO MACHINE} code \citep{kornilov1983}. This computer program  was designed for the population synthesis of the evolution of isolated close binary stars. It was described in detail by \citet{lipunov1996,lipunov2009}, also \citet{cherepashchuk2019} recently depicted specific parameters of the evolutionary scenario that were important for the study of the possible evolution of X-ray novae (with black holes) under the influence of the increased angular momentum loss via magnetic stellar wind of the optical star. So, here (to avoid extensive descriptions that had already been made) we directly referred to corresponding equations in the paper by \citet{cherepashchuk2019} and described parameters of the magnetic field of neutron stars that were important specifically for this manuscript.

The Salpeter initial mass function, the flat initial distribution of binaries on the semi-major axis, and the equiprobable initial mass ratio were accepted. As free parameters of the evolutionary scenario we treated following values: the magnetic field decay time of neutron stars $\tau_{NS}$ (see below), the kick velocity parameter $v_0$ (Equation A8 by \citealp{cherepashchuk2019}), the common envelope efficiency $\alpha_\textrm{CE}$ (Equation A3 by \citealp{cherepashchuk2019}), coefficients of the stellar wind intensity of optical stars $\alpha$ and $\alpha_{WR}$ (Equations A1 and A2 by \citealp{cherepashchuk2019}), the angular momentum loss via the magnetic stellar wind was specified by the $\lambda_\textrm{MSW}$ parameter (Equations A4 and A5 by \citealp{cherepashchuk2019}).

The initial magnetic dipole moment $\mu$ of neutron stars was distributed as

\begin{equation}
f(\log \mu)\propto \textrm{const},
\label{mu}
\end{equation}

\noindent in the range $10^{28}\leq \mu\leq 10^{32}$ G cm$^{3}$, initial rotational periods of neutron stars were $10$ ms. Following formula for the decay of the magnetic field $B$ with time $t$ was used \citep{lipunov2009}:

\begin{equation}
B=\left\{\begin{array}{l}
B_0 \exp(-t/\tau_\textrm{NS}), t < \tau_\textrm{NS}\textrm{ln} (B_0/B_{\textrm{min}}), \\
B_{\textrm{min}}, t\ge \tau_\textrm{NS}\ln (B_0/B_{\textrm{min}}),
\label{field}
\end{array}\right. \end{equation}

\noindent where $B_\textrm{min}=8\times 10^7$ G, $\tau_\textrm{NS}$ is the characteristic decay time of the magnetic field, $B_0$ should be computed from the initial magnetic dipole moment. The Shvartsman-Lipunov classification scheme of states of neutron stars was used to depict them as magnetic rotators \citep{shvartsman1970,shvartsman1971,lipunov1992,lipunov2021}.

\section{Population synthesis}
\label{sec:pop-synth}

As the X-ray nova in this study we accepted the close binary system consisting of the neutron star accretor and the low mass optical star, the mass of the optical star was in the range $0.2-1.1 M_{\odot}$. We computed separately quantities of such binaries (in our galaxy) with the main sequence star and with the star that filled its Roche lobe. Also we computed separately these quantities for systems with orbital periods $\lesssim 0.5$ d and with the magnetic stellar wind timescale $\lesssim 10^8$ yrs (the same subdivisions as in paper by \citealp{cherepashchuk2019}).

Observational limits on the quantity of X-ray novae in the Galaxy were $\approx 300 - 3000$ \citep{chen1997,cherepashchuk2013}, and we used them to find appropriate values of evolutionary parameters to model the quantity of systems under investigation in our calculations. Also as an observational constraint one can treat the fact that among several tens of studied X-ray novae \citep{cherepashchuk2013} there was at least one system with the decrease of the orbital period with much faster rate than it was expected \citep{ponti2017}.

We used following values of free parameters during the modelling of the population of X-ray novae (with neutron stars): $\tau_{NS}$ was taken to be $10^7$, $5\times 10^7$ and $10^8$ yrs, we used zero natal kick velocity ($v_0=0$) and $v_0=180$ km s$^{-1}$, $\alpha_\textrm{CE}$ took values $0.1$, $0.3$, $0.5$, $0.7$, $1.0$, $2.0$, stellar wind coefficients $\alpha=\alpha_{WR}$ were 0.1, 0.3 and 0.5, $\lambda_\textrm{MSW}$ was varied between 1.0 and 0.01. For each set of evolutionary parameters in this study we performed calculations of $10^5$ evolutionary tracks.

Calculations showed that models with the moderately high kick velocity (with $v_0=180$ km s$^{-1}$) practically did not give any significant amount of systems under investigation, so we used zero natal kick velocity to model the population of X-ray novae with neutron stars (in disagreement with the work by \citealp{lipunov1996b} who computed most appropriate values of evolutionary parameters for the population synthesis of binaries using {\ttfamily SCENARIO MACHINE}). This fact was natural, because in this paper we studied the evolution of neutron star binaries with optical companions within very low mass range. Even a small kick in such binary stars can lead to the disruption of the system.

Relatively long time scale of the decay of the magnetic field of neutron stars ($\tau_{NS}=5\times 10^8$ and $10^8$ yrs) did not allow to form required systems (the neutron star in its accreting state paired with the appropriate optical star) with the magnetic stellar wind time $\tau_\textrm{MSW} \lesssim 10^8$ yrs for almost the whole range of $\lambda_\textrm{MSW}$. This result contradicted with observational limits on the quantities of systems under investigation (at least one such system was found), that was why we used only $\tau_{NS}=10^7$ yrs for subsequent computations.

The stellar wind intensity was a bit less important in comparison to our calculations for X-ray novae with black holes \citep{cherepashchuk2019}, because progenitors of neutron stars were less massive in comparison to progenitors of black holes. Nevertheless, values $\alpha=\alpha_{WR}\gtrsim 0.5$ were definitely inappropriate, they did not allow to get any form of studied systems. For subsequent calculations we used $\alpha=\alpha_{WR}=0.3$ that was appropriate for the solar metallicity.

\begin{figure}
\includegraphics[width=\columnwidth]{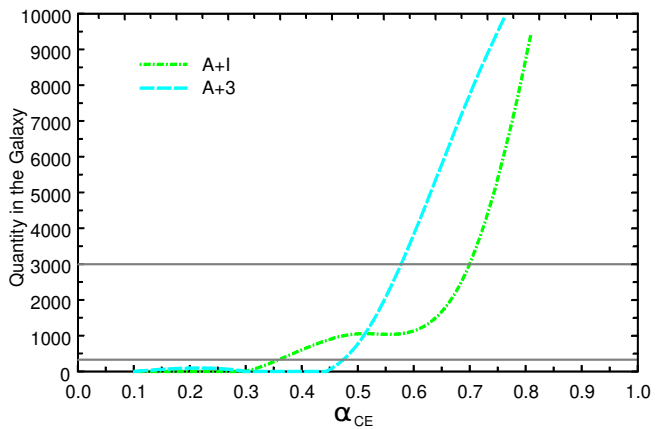}
\vspace{5pt} \caption{The dependence of the quantity in the Galaxy of accreting neutron stars in close pairs with low mass ($0.2-1.1M_{\odot}$) non-degenerate stars (main sequence, ``A+I'', Roche lobe filling, ``A+3'') on the common envelope efficiency $\alpha_\textrm{CE}$, orbital periods were $\lesssim 0.5$ d. The grey bar showed the observational range of possible quantity of X-ray novae in the Milky Way. Values of other evolutionary parameters were: $\tau_{NS}=10^7$ yrs, $v_0=0$, $\alpha=\alpha_{WR}=0.3$, $\lambda_\textrm{MSW}=1.0$.}
\label{c-f-18}
\end{figure}

The common envelope stage was very important, because orbital separations of components in X-ray novae were less or comparable to sizes of main sequence progenitors of neutron stars. So, all X-ray novae should pass this stage, they should be formed from much wider binaries, the common envelope should bring the components close to each other. In Figure \ref{c-f-18} we showed the quantity of studied systems (with orbital periods $\lesssim 0.5$ d) in our galaxy in dependence of the $\alpha_\textrm{CE}$ parameter. As followed from this plot the adequate value of the common envelope efficiency was $\approx 0.5-6$ to satisfy the observational limits (in a good agreement with results obtained by \citealp{lipunov1996b}). For calculations in Figure \ref{c-f-18} the magnetic stellar wind parameter was $\lambda_\textrm{MSW}=1.0$, this value was usually used to describe this kind of wind in most of binaries \citep{tutukov1988}. It also can be noted that if we considered the less intensive stellar wind ($\alpha=\alpha_{WR}=0.1$) the appropriate value of $\alpha_\textrm{CE}$ was in the range $1-1.5$, this fact slightly change several concrete values of evolutionary parameters, but qualitatively all results were the same as for $\alpha=\alpha_{WR}=0.3$. That was why for subsequent calculations we used $\alpha_\textrm{CE}=0.5$. It should be noted that during this stage of our research we tried to roughly describe the whole population of X-ray novae with neutron stars. The next stage was aimed to the most specific and rare systems like AX~J1745.6-2901 with the most rapid orbital angular momentum losses among all X-ray novae; the system should become close, only then it can be significantly affected by the magnetic stellar wind.

\begin{figure}
\includegraphics[width=\columnwidth]{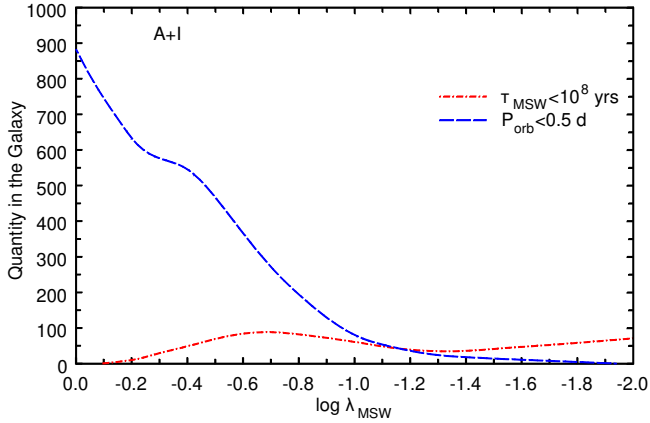}
\vspace{5pt} \caption{The dependence of the quantity of ``A+I'' systems (accreting neutron stars paired with main sequence stars that did not fill their Roche lobes) on the strength of the magnetic field (the coefficient $\lambda_\textrm{MSW}$ was varied from 1.0 to 0.01) for systems with orbital periods $\lesssim 0.5$~d and for systems with the formal value of the magnetic stellar wind time $\lesssim 10^8$ yrs. With the decrease of $\lambda_\textrm{MSW}$ (i.e., with the increase of the angular momentum loss due to the magnetic stellar wind) the quantity of studied systems with period $\lesssim 0.5$ d dropped, because binaries with such orbital periods lived shorter and shorter. The quantity of system with the merger time due to the angular momentum loss in magnetic stellar wind $\lesssim 10^8$ yrs grew, because wider system corresponded to this definition. Observational constraint required at least one rapidly evolving X-ray nova among several tens of X-ray novae, this yielded $\lambda_\textrm{MSW}=0.4-0.15$.}
\label{c-f-19}
\end{figure}

\begin{figure}
\includegraphics[width=\columnwidth]{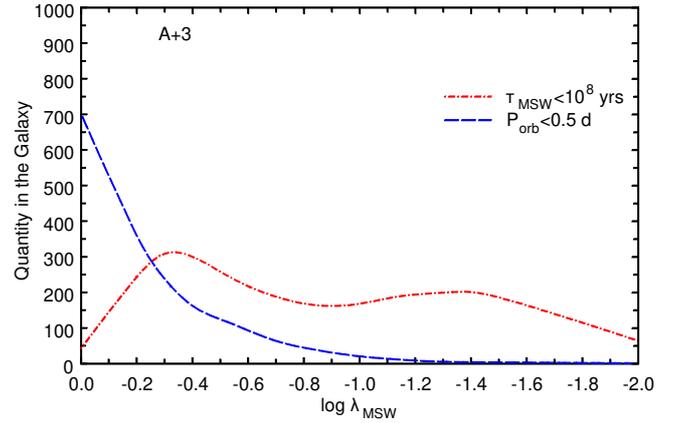}
\vspace{5pt} \caption{The same as Figure \ref{c-f-19} for ``A+3'' systems (accreting neutron stars paired with non-degenerate stars that filled their Roche lobes). The observational constraint required $\lambda_\textrm{MSW}=0.8-0.6$.}
\label{c-f-20}
\end{figure}

In Figures \ref{c-f-19} and \ref{c-f-20} we showed dependencies of quantities of studied systems on the magnetic stellar wind parameter $\lambda_\textrm{MSW}$, separately for systems ``neutron star + main sequence star'' and ``neutron star + optical Roche lobe filling star''. $\lambda_\textrm{MSW}$ was varied in the range 1.0-0.01, $\lambda_\textrm{MSW}=1.0$ corresponded to the usual value of the magnetic wind intensity. The lower this coefficient, the stronger the influence of the magnetic wind on the semi-major axis of the system (the shorter the potential merger time). The quantity of close pairs with orbital periods $\lesssim 0.5$ d dropped with the decrease of $\lambda_\textrm{MSW}$, because the merger time (and, therefore, the lifetime as the X-ray nova) became less. On the other hand the quantity of studied systems with the magnetic stellar wind time $\lesssim 10^8$ increased with the decrease of $\lambda_\textrm{MSW}$, because wider systems satisfied the selection criterion.

The observational limit consisted in that there was at least one X-ray nova with evidently fast decrease of the orbital period among several tens of X-ray novae. As can be seen from Figures \ref{c-f-19} and \ref{c-f-20} this requirement can be met for $\lambda_\textrm{MSW}=0.4-0.15$ in the case of ``neutron star + main sequence star'' binaries and $\lambda_\textrm{MSW}=0.8-0.6$ in the case of ``neutron star + optical Roche lobe filling star'' systems.

A possible evolutionary track of the system that include the stage ``neutron star + main sequence star'' with the merger time $\lesssim 10^8$ years was presented in Table \ref{tr-fast}, it was calculated using $\lambda_\textrm{MSW}=0.1$. The track with the same initial conditions and $\lambda_\textrm{MSW}=1.0$ (ordinary magnetic wind) we presented in Table \ref{tr-ordinary}. Values of other evolutionary parameters were the same for both tracks. These tables showed the influence of the increased angular momentum loss due to the magnetic stellar wind. The evolution of stars in both tracks was the same before the common envelope. For $\lambda_\textrm{MSW}=1.0$ the magnetic wind played no significant role in the system's evolution (Table \ref{tr-ordinary}) also after the common envelope. For $\lambda_\textrm{MSW}=0.1$ the secondary star started to emit the strong magnetic stellar wind, this wind took away a significant amount of the system's angular momentum. The growth of the semi-major axis due to the high mass loss by the primary star (Wolf-Rayet star) was slowed down. It allowed the Wolf-Rayet star to fill its Roche lobe (BB stage). Then the separation of components decreased due to the magnetic wind and the system became much closer before the supernova explosion in comparison to the track with $\lambda_\textrm{MSW}=1.0$. After the supernova explosion the semi-major axis in case of $\lambda_\textrm{MSW}=1.0$ stayed practically unchanged. The only evolution in the system was the evolution of the neutron star from the radio pulsar state to the accretor state (through ejecting and propeller states) up to the age of the Universe. In case of $\lambda_\textrm{MSW}=0.1$ the second companion filled its Roche lobe (because of the angular momentum loss due to the magnetic wind) and (about one thousand years from the supernova explosion) the system became ``neutron star + optical star filling its Roche lobe'' with characteristic lifetime about or less than 20 million yrs.

\begin{table*}
\centering
\caption{A sample evolutionary track that included the stage of a close binary system consisting of an accreting neutron star and a
low-mass non-degenerate star. The magnetic stellar wind was strong ($\lambda_\textrm{MSW}=0.1$). Columns depicted following
values: ``System'' was the composition of the system, $\Delta T$ was the duration of the stage, $M_1$ was the mass of the primary (initially more
massive) star, $M_2$ was the mass of the secondary star (both $M_1$ and $M_2$ were in solar masses), $a$ was the semi-major axis in solar radii, $P_{orb}$ was the orbital period in days, $e$ was the eccentricity, $T$ was the time since the beginning of the evolution (both $\Delta T$ and $T$ were in millions of years). Marks of evolutionary stages were: ``I'' was the main-sequence star, ``II'' was the super-giant, ``3'' was the star filling its Roche lobe, ``WR'' was the
Wolf-Rayet star, ``BB'' was the Wolf-Rayet star filling its Roche Lobe, ``CE'' was the common envelope, ``MSW'' indicated the strong angular momentum loss in the magnetic stellar wind, ``Psr'' was the radio pulsar, ``Ej'' was the ejecting neutron star the did not show itself as the radio pulsar due to the free-free absorption of its radio emission in the wind of the companion and in the interstellar medium (see the nomenclature of compact magnetized objects by \citealp{lipunov1992}), ``P'' was the propeller neutron star, ``A'' was the neutron star accretor, for the ``Final state'' see the text. We used following values of scenario parameters: $\alpha_\textrm{CE}=0.5$, $\alpha=\alpha_{WR}=0.3$, $v_0=0$, $\tau_{NS}=10^7$ yrs.}
\label{tr-fast}
\begin{tabular}{@{}cccccccc@{}}
\hline
System & $\Delta T$ & $M_1$ & $M_2$ & $P_{orb}$ & $a$ & $e$ & $T$ \\
\hline
I+I & 16 & 12.30 & 0.58 & 2875 & 2200 & 0 & 0.0 \\
I+I & & 11.86 & 0.58 & 2926 & 2200 & 0 & 16 \\
II+I & 1.6 & 11.86 & 0.58 & 2926 & 2200 & 0 & 16 \\
II+I & & 11.56 & 0.58 & 3166 & 2300 & 0 & 17 \\
3+I, CE & 0.01 & 11.56 & 0.58 & 3166 & 2300 & 0 & 17 \\
3+I, CE & 0.01 &  3.36 & 0.58 & 1 & 7.3 & 0 & 17 \\
WR+I, MSW & 0.94 &  3.36 & 0.58 & 1 & 7.3 & 0 & 17 \\
WR+I, MSW & 0.94 &  2.35 & 0.58 & 1.8 & 9.8 & 0 & 18 \\
BB+I, MSW & 0.01 &  2.35 & 0.58 & 1.8 & 9.8 & 0 & 18 \\
BB+I, MSW & 0.01 &  1.92 & 0.58 & 0.08 & 1.2 & 0 & 18 \\
\multicolumn{8}{c}{SN Ib} \\
Psr+I, MSW & 0.0001 &  1.30 & 0.58 & 0.18 & 1.8 & 0.34 & 18\\
Ej+I, MSW & &  1.30 & 0.58 & 0.18 & 1.8 & 0.34 & 18 \\
Ej+3, MSW & 0.0003 & 1.30 & 0.58 & 0.18 & 1.8 & 0.34 & 18 \\
Ej+3, MSW & & 1.30 & 0.58 & 0.18 & 1.8 & 0.34 & 18 \\
P+3, MSW & 0.0005 &  1.30 & 0.58 & 0.18 & 1.8 & 0.34 & 18 \\
P+3, MSW & &  1.30 & 0.58 & 0.18 & 1.8 & 0.34 & 18 \\
A+3, MSW & 2 &  1.30 & 0.58 & 0.18 & 1.8 & 0.34 & 18 \\
\multicolumn{7}{c}{Final state} & $\gtrsim 20$ \\
\end{tabular}
\end{table*}
 
\begin{table*}
\centering
\caption{The same as Table \ref{tr-fast} for $\lambda_\textrm{MSW}=1$. Additional notations depicted following states of the neutron star: ``GP'' was the geopropeller and ``P'' was the georotator.}
\label{tr-ordinary}
\begin{tabular}{@{}cccccccc@{}}
\hline
System & $\Delta T$ & $M_1$ & $M_2$ & $P_{orb}$ & $a$ & $e$ & $T$ \\
\hline
I+I & 16 & 12.30 & 0.58 & 2875 & 2200 & 0 & 0 \\
I+I & & 11.86 & 0.58 & 2926 & 2200 & 0 & 16 \\
II+I & 1.6 & 11.86 & 0.58 & 2926 & 2200 & 0 & 16 \\
II+I & & 11.56 & 0.58 & 3166 & 2300 & 0 & 17 \\
3+I, CE & 0.01 & 11.56 & 0.58 & 3166 & 2300 & 0 & 17 \\
3+I, CE & & 3.36 & 0.58 & 1 & 7.3 & 0 & 17 \\
WR+I & 0.94 & 3.36 & 0.58 & 1 & 7.3 & 0 & 17 \\
WR+I & & 2.35 & 0.58 & 1.8 & 9.8 & 0 & 18 \\
\multicolumn{8}{c}{SN Ib} \\
Psr+I & 0.22 & 1.30 & 0.58 & 8 & 23 & 0.57 & 18 \\
Ej+I & & 1.30 & 0.58 & 8 & 23 & 0.57 & 18 \\
GP+I & 1 & 1.30 & 0.58 & 8 & 23 & 0.57 & 18 \\
GP+I & & 1.30 & 0.58 & 8 & 23 & 0.57 & 20 \\
G+I & 100 & 1.30 & 0.58 & 8 & 23 & 0.57 & 20 \\
G+I & & 1.30 & 0.58 & 8 & 23 & 0.57 & 120 \\
A+I & 15000 & 1.30 & 0.58 & 8 & 23 & 0.57 & 120 \\
A+I & & 1.30 & 0.58 & 8 & 20 & 0.48 & $1.5\times 10^4$ \\
\end{tabular}
\end{table*}

The final result of the track in Table \ref{tr-fast} was debateable. If the core of the mass losing optical star was not enriched with helium at the time when the mass value became equal to $\approx 0.3 M_{\odot}$ the magnetic stellar wind potentially can stop \citep{tutukov2018}, because the convection from the envelope penetrates the radiative core and causes the decay of the magnetic field. Also this star can lose its magnetic stellar wind when its mass became equal to the brown dwarf mass or even a giant planet \citep{paczynski1981}. In these cases the subsequent evolution should be similar to the track in Table \ref{tr-ordinary}. The possibility of the unlimited approach of the optical star to the neutron star also remains. In this case the system potentially can form a short living low mass Thorne-Zytkow object.

\section{Magnetic field estimations}
\label{magnetic-estimations}

In this section, we perform a rough estimation for the magnetic fields of the optical stars that can produce the observed orbital-period derivatives of AX J1745.6-2901, KV UMa, A0620-00, and Nova Muscae 1991 (systems in Table \ref{tr-fast} of this paper and in Table A2 of \citealp{cherepashchuk2019} are also considered). Two different equations are used to derive the magnetic fields, the first one is Equation (\ref{induction}), and the second one is Equation 4 of \citet{chen2019}.

Equation (\ref{induction}) was derived in supposition of the frozen magnetic field, it led to $\dot P\sim B_\textrm{d}^2$. Equation (4) by \citet{chen2019} was derived for the dipole magnetic field of the optical star by \citet{justham2006}, it led to $\dot P\sim B_\textrm{d}$. The first case ($\dot P\sim B_\textrm{d}^2$) seems to be useful to explain the empirical Skumanich law \citep{skumanich2019}, also the supposition of the frozen magnetic field makes it possible to avoid the discussion of the field structure. The second case also can be used for estimations, however the structure of the field of the optical star at least in case of Ap stars can be a matter of debates, the field can be a superposition of collinear, centered, dipole, quadrupole, and octupole components  \citep{landstreet2000,mathys2017}.

\subsection*{AX~J1745.6-2901}

The orbital period is $P_\textrm{orb}=30 063.74\pm 0.14$ s \citep{hyodo2009}, its derivative is $\dot P_\textrm{orb}=-4.03\pm 0.32 \times 10^{-11}$ s s$^{-1}$ \citep{ponti2017}, the masses of both components are uncertain, here we assume $M_\textrm{a}=1.4M_\odot$ for the neutron star and $M_\textrm{d}=0.8 M_\odot$ for the optical donor following reasons from the paper by \citep{ponti2017}, the radius of the donor star equals to the radius of its Roche lobe $R_\textrm{d}=0.88R_\odot$ (this value can be estimated from the Kepler's third law and assumed masses of the stars in the system), the wind's velocity is taken to be equal to $w=2v_\textrm{esc}$ (taking into account values for massive stars, see, e.g., \citealp{dejager1980}), the escape velocity is $v_\textrm{esc}=5.88\times 10^7$ cm s$^{-1}$. These values and Equation \ref{induction} give the induction value $B_\textrm{d}\approx 400$ G.

For estimations using Equation 4 by \citet{chen2019} there is a need for several other values of system parameters in addition to quantities above. The X-ray luminosity in the low state is $L_\textrm{x}=2\times 10^{35}$ erg s$^{-1}$ \citep{maeda1996}, it gives the accretion rate $\dot M_\textrm{a}=3\times 10^{-11}M_\odot$ yr$^{-1}$. The quantity of the accretion rate onto the neutron star can be calculated, for example, using equation \citep{lipunov1992}

\begin{equation}
L_\textrm{x}=\frac{\dot M_\textrm{a}GM_\textrm{a}}{2R_\textrm{in}},
\label{xray}
\end{equation}

\noindent where the radius of the inner border of the accretion disc $R_\textrm{in}$ equals to the radius of the neutron star $\approx 10^6$ cm. The semi-major axis of the system can be estimated using Kepler's third law with masses assumed above, so $a\approx 3R_\odot$. The arbitrary parameter $f$ is taken to be equal to 0.001 (as in the paper by \citealp{chen2019}) for all systems under consideration. The subsequent estimation of the induction of the magnetic field is $B_\textrm{d}\approx 1700$ G.

\subsection*{KV UMa}

According to \citet{g-h-2012ApJ} the orbital period is $P_\textrm{orb}=0.16993404\pm 0.00000005$ d, its derivative is $\dot P_\textrm{orb}=-(5.8\pm 2.1)\times 10^{-11}$ s s$^{-1}$. The mass of the black hole is $M_\textrm{a}=(7.06-7.24)M_\odot$ \citep{cherepashchuk2019}, the mass of the optical star is less certain and can be $M_\textrm{d}=0.20 \pm 0.02 M_\odot$ \citep{g-h-2012ApJ} and $M_\textrm{d}=0.10 \pm 0.01 M_\odot$ \citep{petrov2017} depending on the mass ratio of the components obtained in these papers \citep{cherepashchuk2019}. For the purpose of this paper we use the first value ($M_\textrm{d}=0.20 \pm 0.02 M_\odot$), and therefore the semi-major axis of the binary and the radius of the donor are assumed to be equal to $2.77R_\odot$ and $0.39R_\odot$ correspondingly, the escape velocity is $v_\textrm{esc}=4.44\times 10^7$ cm s$^{-1}$. Equation (\ref{induction}) gives $B_\textrm{d}\approx 1500$ G. \citet{chen2019} for $B_\textrm{d}=5000$ G computed using their Equation (4) $\dot P_\textrm{orb}=-5.3\times 10^{-11}$ s s$^{-1}$ that is within the errors of the observed value.

\subsection*{A0620-00}

The orbital period is $P_\textrm{orb}=0.323 014\pm 0.000004$ d \citep{mcclintock1986}, its derivative is $\dot P_\textrm{orb}=-1.90\pm 0.26\times 10^{-11}$ s s$^{-1}$ \citep{g-h-2014}. The masses of the components are $M_\textrm{a}=6.61^{+0.23}_{-0.17} M_\odot$ and $M_\textrm{d}=0.40 \pm 0.01 M_\odot$ \citep{g-h-2014}. The semi-major axis of the system and the radius of the donor are assumed to be equal to $4.17R_\odot$ and $0.74R_\odot$ correspondingly, the escape velocity is $v_\textrm{esc}=4.55\times 10^7$ cm s$^{-1}$. Equation (\ref{induction}) gives $B_\textrm{d}\approx 400$ G. \citet{chen2019} for $B_\textrm{d}=5000$ G computed using their Equation (4) $\dot P_\textrm{orb}=-6.1\times 10^{-11}$ s s$^{-1}$ that is several times greater than the observed value. 

\subsection*{Nova Muscae 1991}

The orbital period is $P_\textrm{orb}=0.432 606\pm 0.000003$ d \citep{orosz1996}, its derivative is $\dot P_\textrm{orb}=-6.56 \pm 4.03 \times 10^{-10} $ s s$^{-1}$ \cite{g-h-2017}. The masses of the components are $M_\textrm{a}=11^{+2.1}_{-1.4}M_\odot$ and $M_\textrm{d}=0.89 \pm 0.18 M_\odot$ respectively \citep{wu2016}. The semi-major axis and radius of the donor are assumed to be equal to $6R_\odot$ and $1.16R_\odot$ correspondingly, the escape velocity is $v_\textrm{esc}=5.37\times 10^7$ cm s$^{-1}$. Equation (\ref{induction}) gives $B_\textrm{d}\approx 1800$ G. \citet{chen2019} for $B_\textrm{d}=5000$ G computed using their Equation (4) $\dot P_\textrm{orb}=-3.5\times 10^{-11}$ s s$^{-1}$ that is about 10 times less than the observed value, i.e., the required field should be about $50\, 000$ G to explain the observed derivative of the orbital period with mention formula.

\subsection*{Computed tracks}

For the track in Table \ref{tr-fast} for the stage ``A+3, MSW'' $R_\textrm{d}=0.56R_\odot$, $v_\textrm{esc}=6.3\times 10^7$ cm s$^{-1}$, $\dot P=-2.47\times 10^{-10}$ s s$^{-1}$. Combining these quantities and quantities from Table \ref{tr-fast} and substituting them into Equation (\ref{induction}) one can obtain the estimate of the required magnetic induction $B_\textrm{d}\approx 1800$ G. For the track in Table A2 by \citet{cherepashchuk2019} for the stage ``BH+3, MSW'' $R_\textrm{d}=0.38R_\odot$, $v_\textrm{esc}=5.89\times 10^7$ cm s$^{-1}$, $\dot P=-1.34\times 10^{-11}$ s s$^{-1}$. Combining these quantities and quantities from Table A2 by  \citet{cherepashchuk2019} and substituting them into Equation (\ref{induction}) one can obtain the estimate of the required magnetic induction $B_\textrm{d}\approx 1100$ G.

\section{Discussion and conclusions}

This paper is dedicated to the investigation of the influence of the increased magnetic field of the optical star after the common envelope interaction.

We successfully reproduced the current observational limit on the part of anomalous neutron star X-ray novae (with rapid decrease of the orbital period) among all neutron star X-ray novae with known parameters (at least one system with accelerated approach of components among several tens of known X-ray novae with orbital periods $\lesssim 0.5$ d), the result can be found in Figures \ref{c-f-19} and \ref{c-f-20} (Figure \ref{c-f-18} makes it possible to choose the value of $\alpha_\textrm{CE}$ for subsequent calculations). The value of $\lambda_\textrm{MSW}$ required for the appropriate acceleration is $0.8-0.6$ for systems with Roche lobe filling optical stars and $0.4-0.15$ for binaries with optical stars that do not fill their Roche lobes. These $\lambda_\textrm{MSW}$ limits mean less increase of the magnetic field in comparison with three black hole X-ray novae with accelerated approach of components (KV UMa, A0620-00, Nova Muscae), where $\lambda_\textrm{MSW}=0.13$ \citep{cherepashchuk2019}.

The estimated values of the magnetic field induction (using Equation (\ref{induction}) in this paper) for all four anomalous X-ray novae are following: $B_\textrm{d}\approx 400$ G (AX~J1745.6-2901), $B_\textrm{d}\approx 1500$ G (KV UMa), $B_\textrm{d}\approx 400$ G (A0620-00), $B_\textrm{d}\approx 1800$ G (Nova Muscae). For AX~J1745.6-2901 the induction estimated using Equation (4) by \citet{chen2019} is $B_\textrm{d}\approx 1500$ G, parameters of accelerated approach of components in KV UMa, A0620-00, Nova Muscae were estimated by \citet{chen2019} for $B_\textrm{d}\approx 5000$. They found $\dot P$ close to observed values in KV UMa and A0620-00, but their $\dot P$ estimate in Nova Muscae was about 10 times less than the observed value (i.e. the required field should be $\approx 50\, 000$ G).

Tables \ref{tr-fast} and \ref{tr-ordinary} describe the difference in the evolution between the track with the standard magnetic stellar wind ($\lambda_\textrm{MSW}=1.0$) and the track with the amplified magnetic field during the common envelope stage ($\lambda_\textrm{MSW}=0.1$). It can be seen that this amplification makes the binary closer before the supernova explosion and potentially can change the final state of the system. In case of ordinary magnetic wind the system can survive in the same state during a lot of billions of years (practically longer than the age of the Universe). The amplified magnetic field makes the angular momentum losses much stronger. It can lead to a debatable final: to the unlimited approach of the components with the formation of Thorne-Zhytkow object, or to the stop of the magnetic stellar wind if the convection penetrates the radiative core of the optical star during the process of mass loss by this star. The required magnetic field induction for the track in Table \ref{tr-fast} is $B_\textrm{d}\approx 1800$ G. For the black hole track in Table A2 by \citet{cherepashchuk2019} it is $B_\textrm{d}\approx 1100$ G.

\citet{van2019} studied how different prescriptions of the magnetic stellar wind can affect the evolution of low mass X-ray binaries with accreting neutron stars. They found that the standard description of the magnetic stellar wind obtained from the Skumanich law ($\lambda_\textrm{MSW}=1.0$ in terms used in this paper) cannot reproduce all observed binaries (``default'' prescription by \citealp{van2019}). ``Convection busted'', ``intermediate'' and ``wind boosted'' scenarios allowed them to partially or practically fully describe studied binaries with greater angular momentum losses via the magnetic stellar wind in comparison with classical values, i.e. their results are compatible with our conclusion about the amplification of the magnetic stellar wind losses after the common envelope phase. However, direct detailed comparisons of results in this paper and in paper by \citep{cherepashchuk2019} with results by \citet{van2019} and by \cite{chen2019} are not possible, because in our calculations there are only initial and final parameters within one evolutionary stage, internal structure of stars and other parameters as functions between these states (within the same stage) are not calculated by {\ttfamily SCENARIO MACHINE} code (Tables \ref{tr-fast} and \ref{tr-ordinary}).

\citet{chen2019} considered several processes that can change the orbital angular momentum of the X-ray low mass binary system. As the strongest magnetic field in the most favourable for the magnetic braking case they took the surface magnetic field 5 kG, the magnetic braking was taken into account in the form by \citet{spruit1983,rappaport1983} and the additional braking due to the increase of the mass loss via the induced stellar wind (which also interacts with the magnetic field of the optical star) in the form by \cite{justham2006}. They came to a conclusion that the magnetic braking can produce the orbital period derivative value that is equal to maximum of 10\% of observed values in Nova Muscae 1991, the optical star in KV UMa indicated as potentially not producing magnetic brake, because its mass is too low and the star should be fully convective. However, the normal star in the KV UMa system is able to still possess a radiative core (if the core during its evolution was enriched by helium) at least until its mass drops to the mass about $0.1M_\odot$ \citep{iben1984,tutukov1985}, therefore it also is able to possess the magnetic stellar wind.

Calculations by \citet{ohlmann2016} showed the growth of the magnetic field strength up to 100 kGs during the common envelope stage. Despite the fact that field strengths in their calculations were for ``diagnostic purposes'', they came to a conclusion that these strengths can be relevant for red luminous novae. The work by \citet{ohlmann2016} considered magnetohydrodynamical simulations of the dynamical spiral-in during the common envelop phase of the system ``$2M_{\odot}$ red giant + $1M_{\odot}$ main sequence star''. The systems considered here and in the work by \citet{cherepashchuk2019} are very significantly different in masses, so the evolution of the magnetic field of the secondary star in them during the common envelope stage should be computed specifically. In our calculations we assumed that the field can grow, and therefore the angular momentum losses via the stellar magnetic wind can also grow far beyond usually used estimations.

In addition, if the donor's core is slightly evolved it can remain radiative for a longer time down to less masses during the accretion process (see, e.g., Figure 2 by \citealp{iben1984}, model ``N6''). It makes possible to keep the magnetic braking on for a longer time. Therefore the increase of the magnetic field in the common envelope as the explanation of the increased rate of the approach of binary components to each other can be valid in all four systems under consideration here and in paper by \citep{cherepashchuk2019},including KV UMa where the mass of the optical star is less than the lower level for the structure ``radiative core + convective envelope'' for the standard chemical composition, but the radiative core can still survive due to the changes in its chemical composion during its evolution, as can be seen in model ``N6'' in Figure 2 by \citealp{iben1984}. However, the disrupted magnetic braking when the core becomes fully convective seem to be confirmed by population synthesis studies and observational statistics (see, e.g., \citealp{politano2006,schreiber2010}), that is why the final fate of the system in Table \ref{tr-fast} is uncertain (the tidal disruption of the star during the fast unlimited approach of the donor to the relativistic object or the disruption of the magnetic braking and further slow evolution under the radiation of graviational waves).

It seems that the mechanism by \citet{applegate1992} and the influence of the third body potentially can be ruled out as explanations of discussed anomalies, because all X-ray novae with the anomalous rate of change of the orbital period (KV UMa, A0620-00 Nova Muscae 1991, AX~J1745.6-2901) have the negative time derivative. Their orbital periods decrease more rapidly than expected, whereas in case of oscillations of the orbital period due to the orbital motion of the center of mass of the binary system around the center of mass of the triple system or due to oscillations of the optical star's shape the change of the orbital period can also be positive, and the observed orbital period can grow too.

The strongest known magnetic field of a non-degenerate star is the magnetic field of HD 215441 (it is an Ap star) that equals to about 34 kG \citep{babcock1960,romanyuk2021} and even higher in some models \citep{landstreet2000}. So, even the highest estimations the required magnetic field in our scenario are of the same order of magnitude as cases that are already known in nature.

Finally we can conclude that the fast decrease of the orbital period of AX~J1745.6-2901 can be explained by the increase the angular momentum loss in the magnetic stellar wind of the optical star, where the magnetic field was strongly amplified in the common envelope before the supernova explosion. This amplification can be less in comparison with X-ray novae with black holes \citep{cherepashchuk2019}. This statement does not contradict other explanations or replace other mechanisms that can lower the angular momentum of the binary star, all of them can work together (they are the gravitational radiation, the non-conservative mass transfer, the interaction with the circumbinary disc, the magnetic stellar wind, including the magnetic stellar wind with amplified angular momentum outflow) with different efficiencies.

In this research we assumed that the field generated during the common envelope stage penetrates the whole star and remains until the decay of the magnetic field when the core of the optical star becames convective. \citet{ohlmann2016} did not make direct estimations of the lifetime of this generated field, indirectly it seems that they suppose the same as we do in this study, because they try to compare the field strength of their output of the common envelope with magnetic field of white dwarfs. The angular momentum loss by cataclysmic variables (post common envelope binaries with white dwarfs) can be adequately described using $\lambda_\textrm{MSW}=1$ \citep{iben1984}, our results for X-ray novae with neutron stars and black holes show the less value of this parameter (see also \citealp{van2019}). Therefore the dependence of the potential field generation and/or a possible difference of the lifetime of the amplified field on the mass of the components of the common envelope binary should be carefully studied along with the structure of this field. \citealp{ohlmann2016} only considered the dipole field, whereas the field structure can be more complicated \citep{landstreet2000,mathys2017}. A potential role of the optical star's rotation with a rate faster than the synchronous rate also can be a matter of a special study for hypothetical binaries that are young enough to keep the faster rotation (see Equaiton \ref{axil-acceleration}).

\section*{Acknowledgements}

The work was supported by the Russian Science Foundation grant \mbox{17-12-01241} (AMC) and by the Scientific and Educational School of
M. V. Lomonosov Moscow State University ``Fundamental and applied space research''. The authors acknowledge support from the M. V. Lomonosov Moscow State University Program of Development (AIB, AMC, KTS).

The authors are grateful to the anonymous referee for very useful notices that helped to improve the paper and to clarify questions that our investigation raises.

\section*{Data availability}

The data underlying this article will be shared on reasonable request to corresponding authors.

\bibliographystyle{mnras}
\bibliography{bogomazov-x-ray-novae-magnetic}

\bsp	
\label{lastpage}
\end{document}